# Sulfur annealing effect for superconductivity in iron chalcogenide compounds

K. Deguchi[1,2], A. Yamashita[1], T. Yamaki[1,2], H. Hara[1,2], S. Demura[1,2], S. J. Denholme[1], M. Fujioka[1], H. Okazaki[1], H. Takeya[1], T. Yamaguchi[1] and Y. Takano[1,2]

[1] *National Institute for Materials Science, 1-2-1, Sengen, Tsukuba, 305-0047, Japan*

[2] *University of Tsukuba, 1-1-1 Tennodai, Tsukuba, 305-8571, Japan*

E-mail: DEGUCHI.Keita@nims.go.jp

Abstract

We discovered a novel annealing method for Fe-chalcogenide superconductors. It was found that sulfur annealing deintercalated excess Fe via formation of $FeS_2$. Due to its specifics, sulfur annealing is applicable when preparing Fe-chalcogenide-based wires or cables.

**Introduction**

Fe-chalcogenide superconductors are promising materials for applications under a high magnetic field, due to a high upper critical field $H_{c2}$ and extremely low anisotropy [1-3]. The $J_c$ of Fe-chalcogenides thin films exhibits a superior high field performance up to 35 T over those of low temperature superconductors [4]. Furthermore, the simple structure which has a lower degree of toxicity than Fe-pnictides simplifies its synthesis and handling. As a result, Fe-chalcogenide superconductors can be considered as potential candidates for applications such as wires, tapes, and thin films.

It is known that superconductivity in Fe-chalcogenides is strongly suppressed by the presence of an excess Fe [5-8]. The excess Fe partially occupies internal sites between the layers and provides electron doping to the conducting layers. This electron doping suppresses the antiferromagnetic wave vector $Q_s = (0.5, 0.5)$, which is found to correlate with Fe-based superconductivity. This means that electron doping caused by excess Fe does not favor superconductivity. However, the excess Fe cannot be totally removed during the synthetic process. The post-annealing as means of suppressing the electron doping is important in order to achieve superconductivity in Fe-chalcogenide.

So far, we have investigated the effect of the excess Fe and found several methods to suppress the electron doping in the conducting layers. Oxidation through air exposure and oxygen annealing compensate the electron doping resulting from the excess Fe [9-12]. These techniques show the reversibility for appearance of superconductivity in Fe-chalcogenides: removal of oxygen by vacuum annealing suppresses the superconductivity, while renewed oxygen-annealing results in its reappearance. The other technique available is deintercalation of excess Fe. This can be achieved by heating in alcoholic beverages, and in organic acid solutions [13-18]. Various other post-annealing methods using $I_2$, Te vapor, and acid solution are also reported [19-23].

However, these methods are not suitable for wire production. The powder-in tube is a common technique to fabricate Fe-chalcogenide wires. It is difficult to perform post-annealing on a sample located inside a tube via outside intervention. To improve the superconducting property of the wires, a novel annealing method is required. Here, we propose sulfur annealing, which is a new route to induce bulk superconductivity in Fe-chalcogenides and suitable for applications.

**Experimental**

Polycrystalline samples of $FeTe_{0.9}Se_{0.1}$ were prepared by a solid-state reaction using powder Fe (99.998 %) and grains of Te (99.9999 %) and Se (99.999 %). The starting materials with a nominal composition of $FeTe_{0.9}Se_{0.1}$ were mixed, pelletized, and put into a quartz tube. The quartz tube was then evacuated by a rotary pump and sealed. The pellet was heated at 700 °C for 10 hours. The obtained mixture was ground, weighed and pelletized into separate pellets with an approximate weight of about 0.100 g each. The pellets were sealed into an evacuated quartz tube and heated at

700 °C for 10 hours. Polycrystalline samples of FeTe$_{0.8}$S$_{0.2}$ and FeTe were also prepared using a solid-state reaction method as described in refs. [24] and [25]. After furnace cooling, the obtained samples were quickly sealed into evacuated quartz tubes with 0.5 g sulfur grains and annealed at either 100, 200 or 300 °C for 2 hours. The temperature dependence of magnetization was measured using a Quantum Design Magnetic Property Measurement System magnetometer down to 2 K after both zero-field-cooling and field-cooling with an applied field of 10 Oe. The electrical resistivity measurements were performed by a standard DC four-terminal method down to 2 K with a current of 1 mA using a Quantum Design Physical Property Measurement System. Powder X-ray diffraction (XRD) patterns were measured using the $2\theta/\theta$ method with Cu-$K\alpha$ radiation by Rigaku Mini Flex II. The polished surface of sample was observed using a scanning electron microscope (SEM). The actual crystal composition was determined by energy dispersive X-ray spectroscopy (EDX).

### Results and discussion

Fig. 1 shows the temperature dependence of magnetic susceptibility for the as-grown FeTe$_{0.9}$Se$_{0.1}$ sample and the FeTe$_{0.9}$Se$_{0.1}$ samples which was sulfur-annealed at 100, 200, and 300 °C. We found that the sulfur-annealed samples show definitive traces of superconductivity, whereas the as-grown sample does not. A diamagnetic signal is observed from the 100 °C annealed sample. For the sample annealed at 200 °C, the diamagnetic signal is strongly enhanced and a superconducting transition at 12.9 K is observed. By contrast, the annealing at 300 °C degrades the superconductivity and impurity phases appear as evidenced by the change of the offset of the magnetization. We also investigated the sulfur annealing for the Fe-chalcogenide compounds FeTe and FeTe$_{0.8}$S$_{0.2}$. Fig. 2 shows temperature dependence of magnetic susceptibility for FeTe and FeTe$_{0.8}$S$_{0.2}$ annealed with sulfur at 200 °C for 2 h. The sulfur-annealed FeTe$_{0.8}$S$_{0.2}$ shows superconductivity whereas sulfur-annealed FeTe does not. These behaviors are very similar to the case of other annealing methods. Thus, sulfur-annealing method can be universally applied to the Fe-chalcogenide superconductors. In the following paragraph, we discuss the sulfur annealing by use of FeTe$_{0.9}$Se$_{0.1}$ sample in detail.

Fig. 3(a) shows the temperature dependence of electrical resistivity below 20 K for the as-grown FeTe$_{0.9}$Se$_{0.1}$ and the annealed FeTe$_{0.9}$Se$_{0.1}$ sample. The as-grown sample has a broad transition and zero resistivity is achieved at 2.0 K. After annealing at 100 °C and 200 °C, the superconducting transition became sharper. The highest $T_c^{onset}$ 13.5 K and $T_c^{zero}$ 11.4 K are obtained for the 200 °C sulfur-annealed sample. The $T_c^{onset}$ of the sample annealed at 300 °C is almost the same as that of the 200 °C annealed sample. However, zero resistivity was observed below 2.3 K. Fig. 3(b) shows the resistivity curves of the samples in the temperature range from 300 to 2 K. Above $T_c$, resistivity of the as-grown sample monotonically increased with decreasing temperature and showed an anomaly around 50 K. This anomaly is due to the antiferromagnetic and structural transition. The anomaly seems to disappear for the 100 and 200 °C annealed sample, indicating that the antiferromagnetic order can be suppressed by sulfur annealing. The resistivity curve of the sample annealed at 200 °C, exhibits a broad hump around 60 K. This behavior is quite similar to that of the oxygen annealed Fe-chalcogenides. The semiconducting-like behavior is observed for the sample annealed at 300 °C. This could be due to a decomposition of the Fe-chalcogenide structure and an increase of impurity phases. Combining with the result of the magnetic susceptibility measurement, we concluded that the optimum temperature of sulfur annealing is 200 °C.

As mentioned in the introduction, the oxidation and the deintercalation of excess Fe induce bulk superconductivity in Fe chalcogenides through different mechanisms. The question then is how sulfur annealing induces superconductivity. To investigate the role of sulfur annealing, XRD and EDX analysis was carried out to compare the as-grown $FeTe_{0.9}Se_{0.1}$ sample with the $FeTe_{0.9}Se_{0.1}$ sample annealed at 200 °C.

Fig. 4 shows the X-ray profiles for the as-grown $FeTe_{0.9}Se_{0.1}$ sample and the $FeTe_{0.9}Se_{0.1}$ sample which underwent sulfur-annealing at 200 °C for 2 hours. The solid circles and squares indicate the impurity phases of $FeTe_2$ and $FeS_2$, respectively. With the exception of these impurity phases, all the other peaks can be characterized by the space group *P*4/*nmm*. For the as-grown sample, a peak belonging to the impurity phase $FeTe_2$ was observed around 32 degrees. By contrast, $FeS_2$ peaks were observed for the 200 °C annealed sample. The calculated lattice constants *a* and *c* of the as-grown sample are 3.8169(5) and 6.2407(9), respectively. Those of the annealed sample are 3.8184(4) and 6.2452(9), respectively, indicating that the sulfur annealing expands the lattice.

Fig. 5 shows SEM images and EDX mappings of the as-grown sample (a) (b) and a specific region of the sample annealed at 200 °C (c) (d). The elemental mapping analysis showed that the as-grown sample is almost homogeneous and that the actual composition is $Fe_{1.08}Te_{0.92}Se_{0.08}$. After sulfur annealing, we found that there are some sulfur-rich regions in the sample. In these regions, the elemental content of Fe and Te is significantly changed. To investigate the change in detail, a line scan near the boundary was carried out. The SEM image and the analyzed positions marked as № 0 to 30 are shown in Fig. 6(a). In this image, we can see the two areas: the grayish (№ 0 to 22) and the darker area (№ 22 to 30). The elemental contents of Fe, Te, Se, and S are summarized in Fig. 6(b). In the darker area, Te content is considerably lower whereas S content is higher. The main composition of this area is $FeS_2$, which is observed in XRD patterns. Interestingly, Fe content in the darker area is higher compared to the grayish area. It indicates that Fe was deintercalated from the grayish area and moved to the darker area. In fact, the average composition of the grayish area is $Fe_{0.98}Te_{0.91}Se_{0.09}$. In comparison to the as-grown sample, Fe content in the grayish area is clearly lower and the decrement of Fe is corresponding to the amount excess Fe. Therefore, we concluded that sulfur annealing deintercalated excess Fe from Fe-chalcogenide phase via formation of $FeS_2$, and hence bulk superconductivity is induced by sulfur-annealing.

Conclusions

We discovered a new annealing method for Fe-chalcogenides. The sulfur annealing deintercalated excess Fe via formation of $FeS_2$ and induced bulk superconductivity. This annealing method does not require a specific environment such as $O_2$ gas, Te vapor, and various solutions. Thus sulfur annealing is applicable for preparing Fe-chalcogenide based wires or cables.

Acknowledgement

This work was partly supported by a Grant-in-Aid for Scientific Research (KAKENHI). This research was partly supported by Strategic International Collaborative Research

**Figure captions**

Fig.1. Temperature dependence of the magnetic susceptibility from 15 to 2 K for the as-grown FeTe$_{0.9}$Se$_{0.1}$ and the sulfur-annealed 100, 200, and 300 °C for 2 hours FeTe$_{0.9}$Se$_{0.1}$.

Fig. 2. Temperature dependence of the magnetic susceptibility from 15 to 2 K for the FeTe and the FeTe$_{0.8}$S$_{0.2}$ samples annealed at 200 °C for 2 hours.

Fig.3. (a) Temperature dependence of the electrical resistivity for the as-grown FeTe$_{0.9}$Se$_{0.1}$ and the sulfur-annealed 100, 200, and 300 °C for 2 hours FeTe$_{0.9}$Se$_{0.1}$ sample. (a) At a temperature near the superconducting transition (b) from 300 to 2 K.

Fig. 4. Powder X-ray diffraction patterns for the as-grown and the sulfur-annealed FeTe$_{0.9}$Se$_{0.1}$ at 200°C. The asterisks indicate the FeTe$_2$ and FeS$_2$ phases.

Fig. 5. SEM image and EDX mapping of FeTe$_{0.9}$Se$_{0.1}$. (a)(b) as-grown sample. (c)(d) sulfur-annealed 200°C for 2 hours sample.

Fig. 6 (a) SEM image of the sulfur-annealed sample. Positions marked № 0 to 30 are analyzed by EDX. (b) Position dependence of elemental content of Fe, Te, Se, and S.

Fig.1

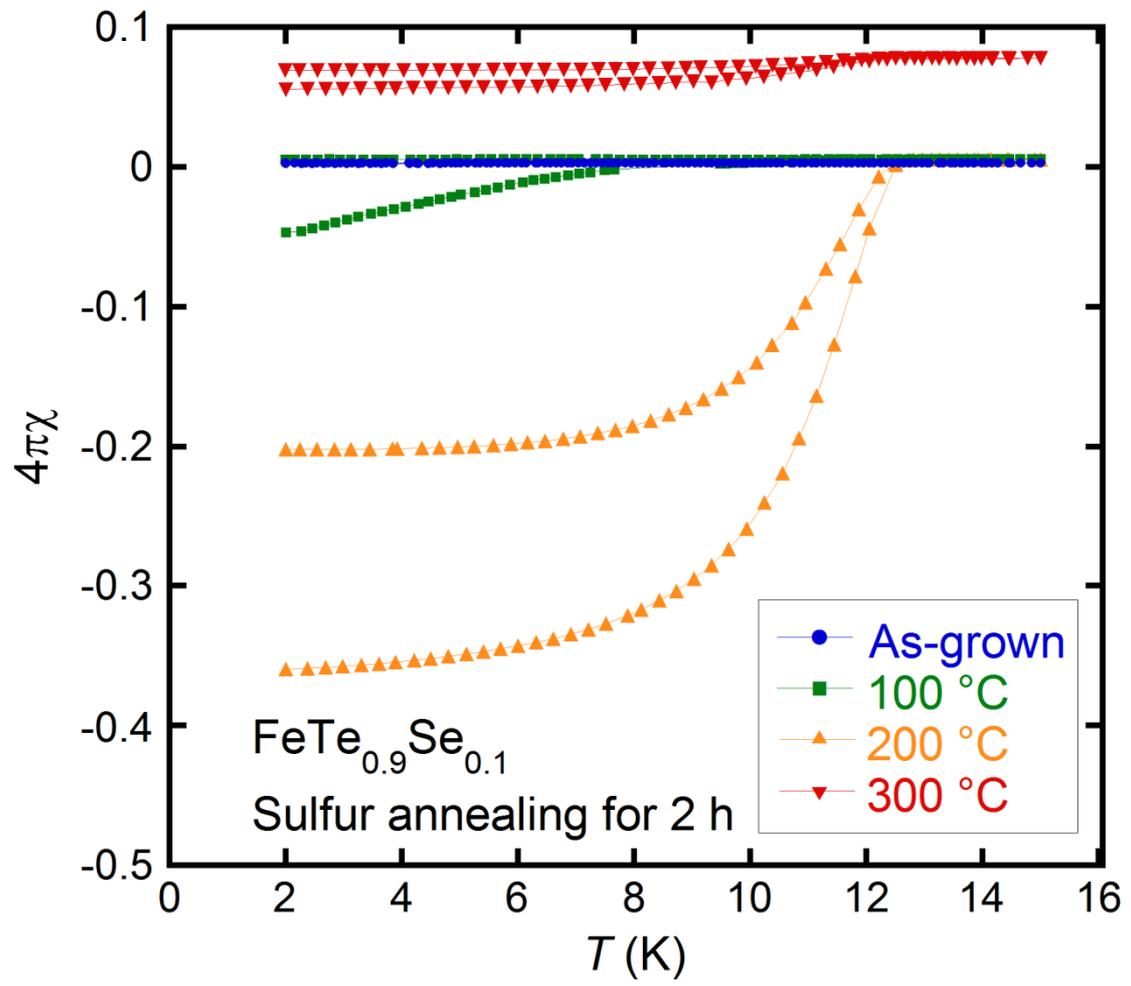

Fig. 2

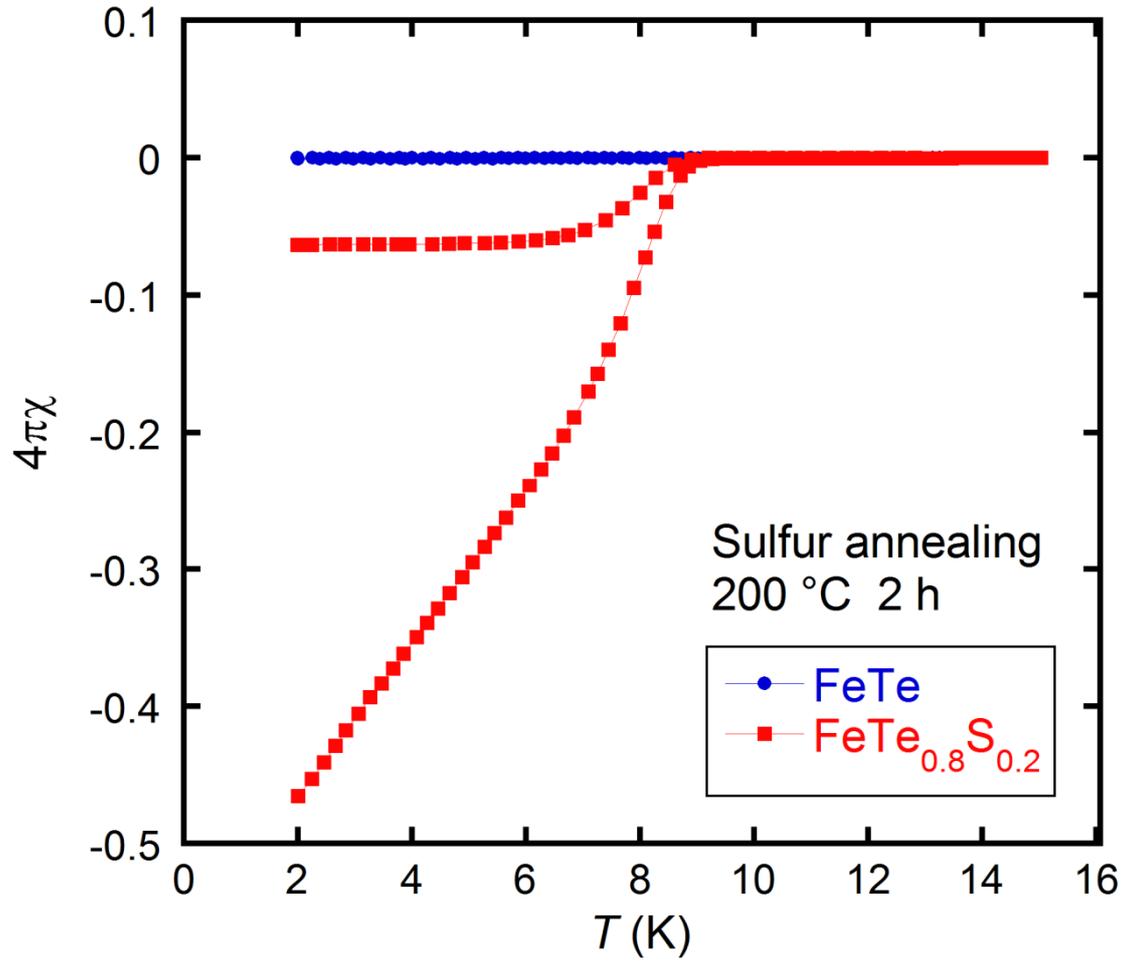

Fig. 3a

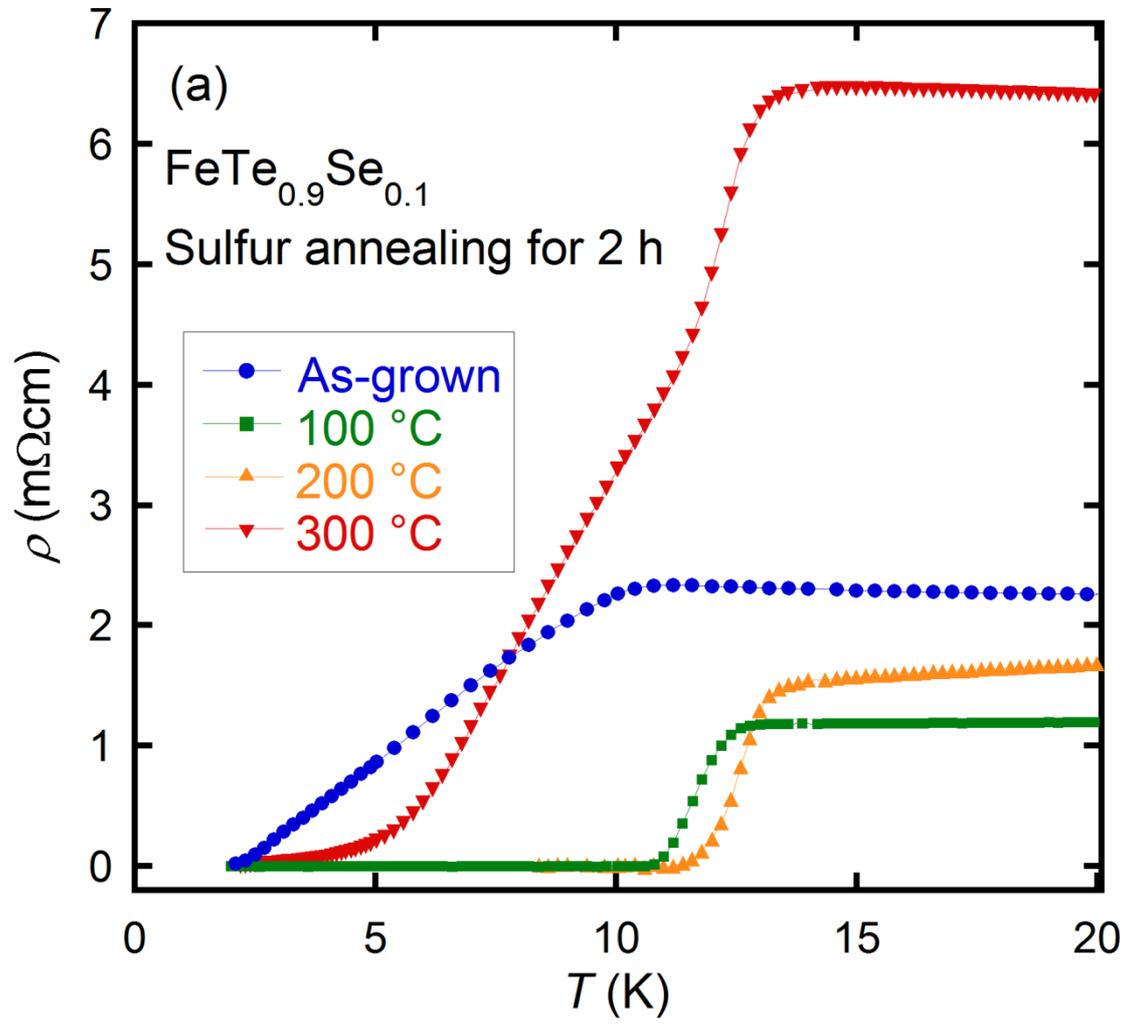

Fig. 3b

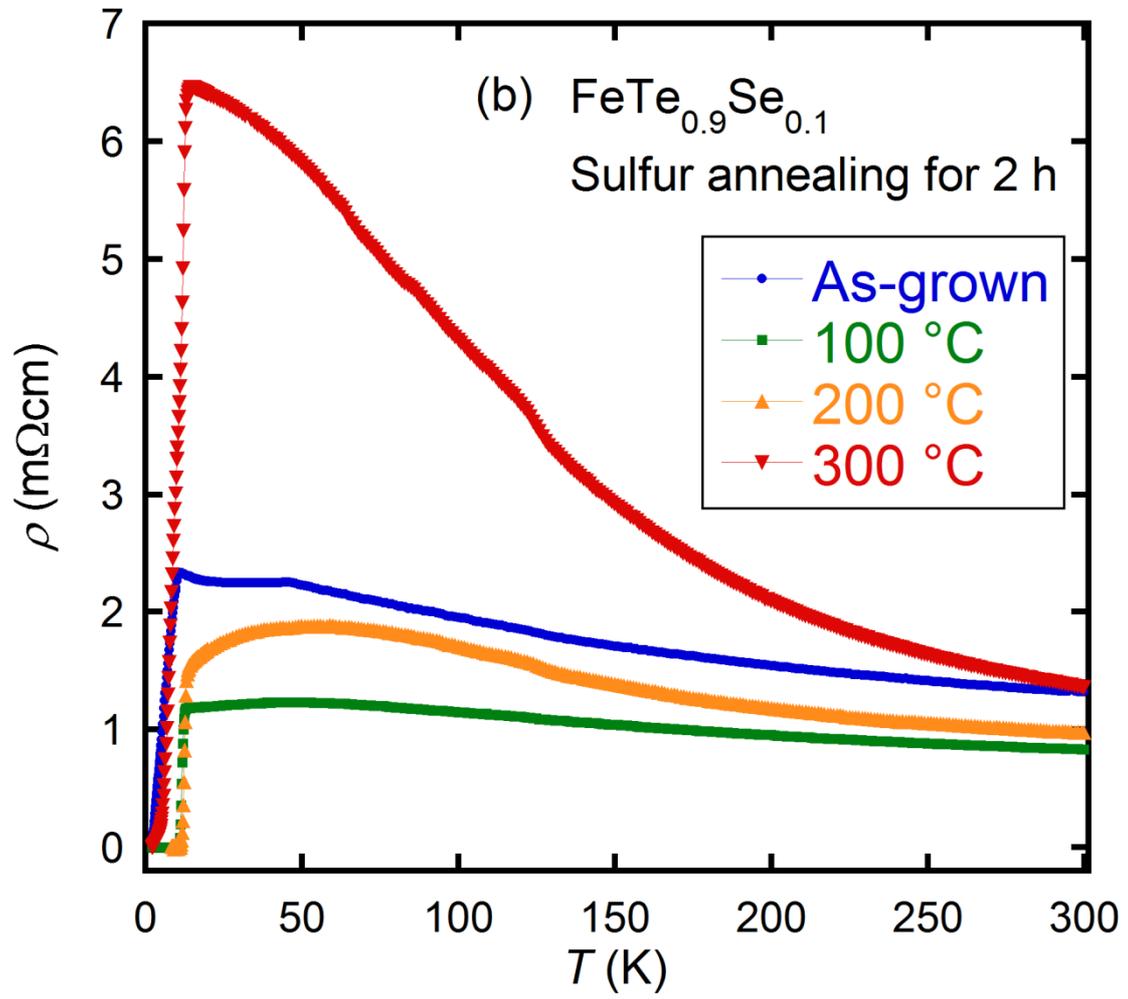

Fig. 4

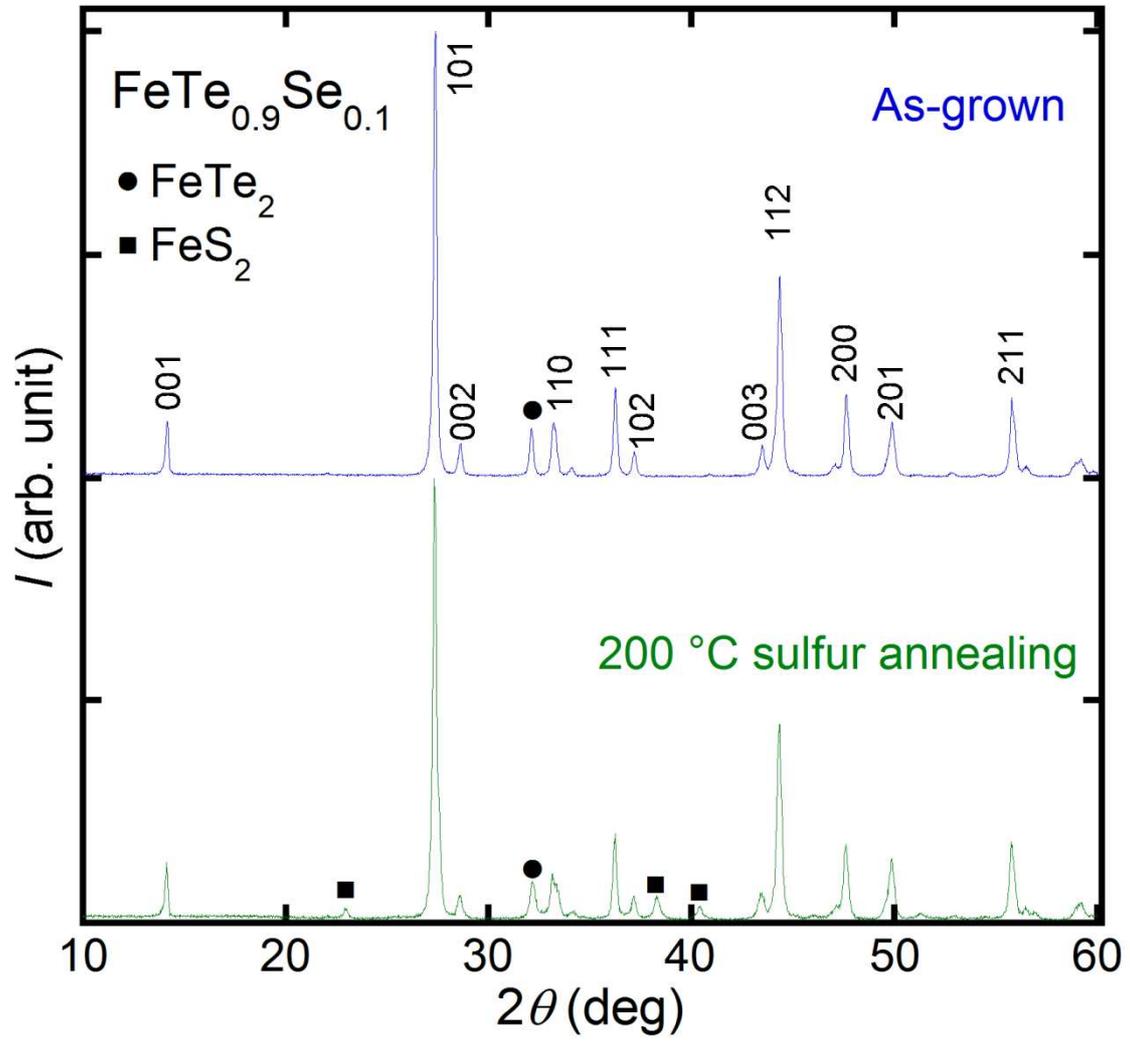

Fig. 5

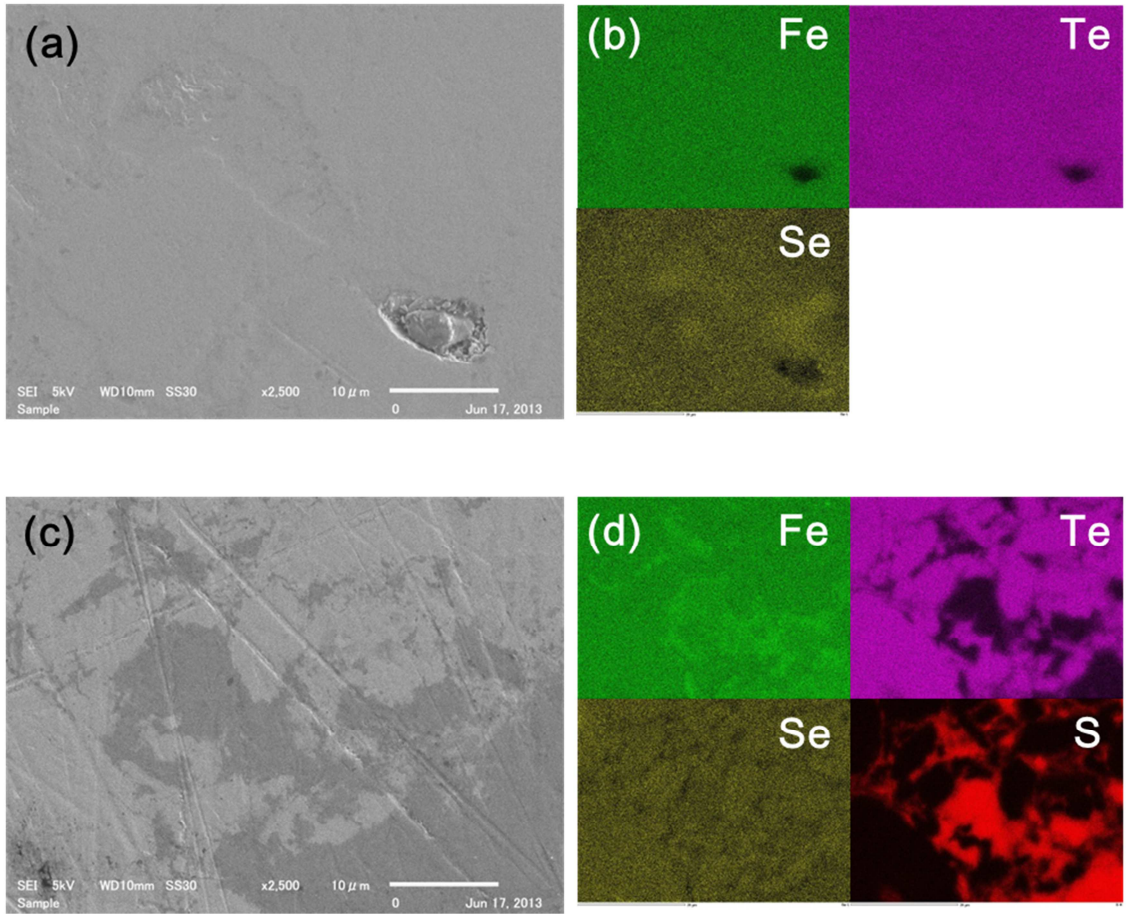

Fig. 6a

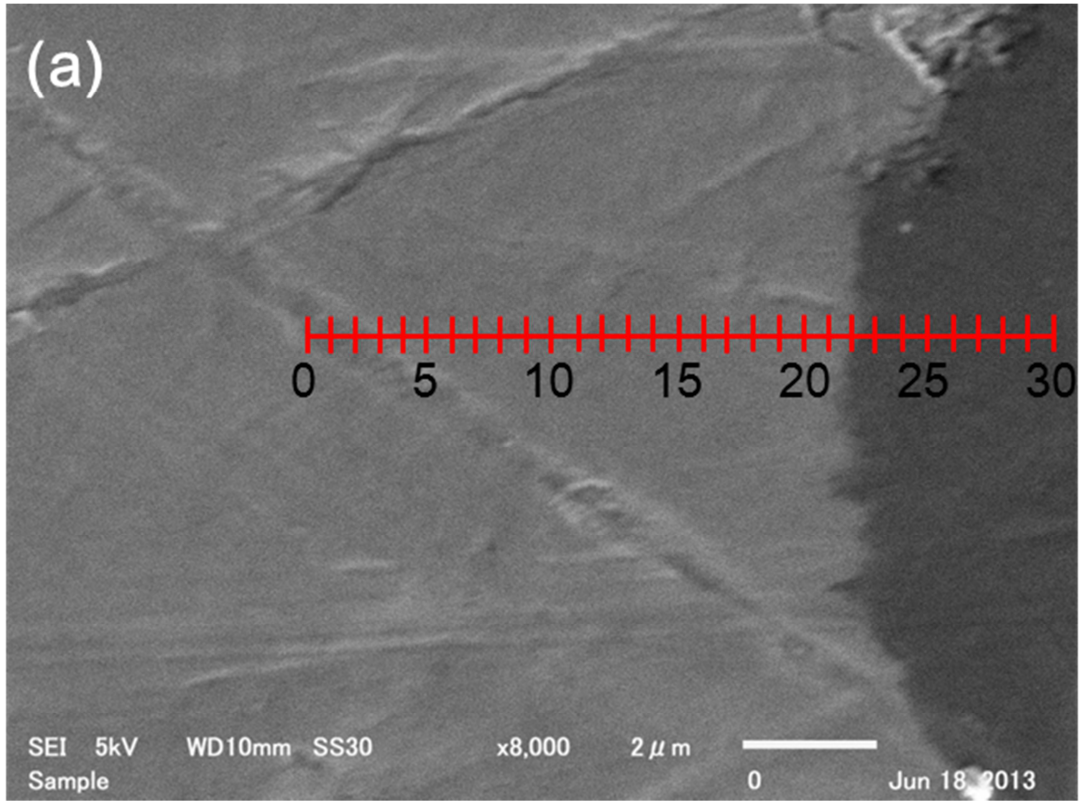

Fig. 6b

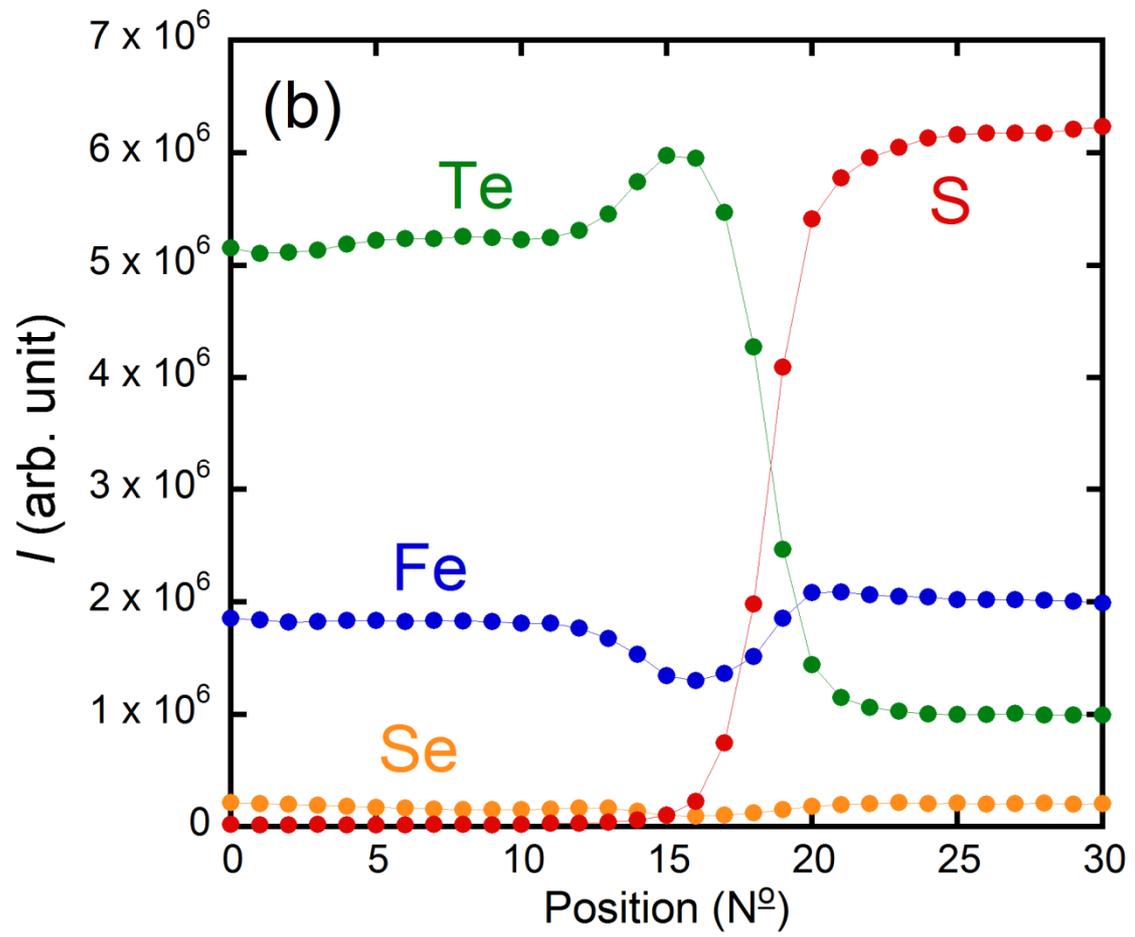